\documentclass[12pt,a4paper]{article}
 \usepackage{graphicx}

\begin{document}
\title{On the dynamics of interacting populations in 
presence of state dependent fluctuations}


\author{Nikolay K. Vitanov$^1$, Zlatinka I. Dimitrova$^2$}
\date{$^1$ Institute of Mechanics, Bulgarian Academy of Sciences,
Acad. G. Bonchev Str., Bl. 4, 1113 Sofia, Bulgaria\\
$^2$ "G. Nadjakov" Institute of Solid State Physics,
Bulgarian Academy of Sciences,
Blvd. Tzarigradsko Chaussee 72, 1784 Sofia, Bulgaria}

\maketitle
\begin{abstract}
We discuss several models of the dynamics of interacting populations.
The models are constructed by nonlinear differential equations and have 
two sets of parameters: growth rates and coefficients of interaction between 
populations. We assume that the parameters depend on the densities of the 
populations. In addition the parameters can be influenced by different factors 
of the environment. This influence is modelled by  noise terms in the equations 
for the growth rates and interaction coefficients. Thus the model differential
equations become stochastic. In some particular cases these equations 
can be reduced to a Foker-Plancnk equation for the probability density
function of the densities of the interacting populations.  
\end{abstract}
\section{Introduction}
In this paper we shall discuss several models of  the dynamics of interacting 
biological populations. Usually such models  consist of nonlinear ordinary differential equations 
for the population densities \cite{murr}-\cite{dv5}. Two sets of parameters
are presented in the models: growth rates and coefficients of interaction between
the populations. The basic assumption in the discussed below models is that 
the model parameters depend on the
densities of the populations. The new point in this study is the assumption that 
the model parameters can depend also on the environment. This influence 
will be modelled by noise terms. Thus the model equations will become nonlinear
stochastic differential equations. The kind of noise will be multiplicative noise
(noise that depends on the populations densities) or more complicated kind of noise.  
\par
The result of the influence of the environment fluctuations is that instead of 
equations for the trajectories of the populations in the phase space of the 
population densities we will have to write and solve equations for the 
probability density functions of the densities of the interacting populations.
\par
Below we shall discuss the models in order of their increasing mathematical 
complexity. We shall start with inclusion of additive noise only in the growth 
rates of populations. This will lead to arising of multiplicative noise in 
the model equations. Then we shall consider a model with additive noise in the 
coefficients of interaction between the populations. The third model will 
contain additive noise in the both sets of parameters: in growth rates and in 
the interaction coefficients. Next we shall consider a model with 
multiplicative state dependent noises in the growth rates and in the 
interaction coefficients of the model equations. Finally we shall show a part 
of methodology for reduction of the nonlinear stochastic differential equations 
to a Fokker-Planck equation for the probability density function of the 
spatial densities of the populations. Several concluding remarks are given at 
the end of the paper.
\section{Model equations without influence of environmental fluctuations}
The classical model of interacting populations is based on a system of 
nonlinear  ordinary differential equations of the Lotka-Volterra kind:
\begin{equation}\label{b1}
\dot{\rho}_i = r_i \rho_i(t) \left( 1- \sum_{j=1}^n \alpha_{ij} \rho_j(t) 
\right).
\end{equation}
in Eqs.(\ref{b1}) $\rho_i$ are the densities of the population members, $r_i$ 
are the growth rates (that can be negative if the number of deaths in the 
corresponding population  is larger than the number of births).  $\alpha_{ij}$ 
are coefficients of interaction between the populations $i$ and $j$. 
$\dot{\rho}_i$ denotes the time derivative of the density $\rho_i$.
\par
Let us now suppose \cite{dv1}-\cite{dv7} that the birth rates and interaction 
coefficients depend on the density of the populations:
\begin{equation}\label{b2}
r_i = r_i^0 \left( 1 + \sum_{j=1}^n r_{ij} \rho_j\right); \ \
\alpha_{ij} = \alpha_{ij}^0 \left(1+ \sum_{j=1}^n \alpha_{ijk} \rho_k \right)
\end{equation} 
In Eq.(\ref{b2}) $r_{ij}$ and $\alpha_{ijk}$ are parameters.
The substitution of Eq.(\ref{b2}) in Eq.(\ref{b1}) leads to a system
of model equations of the kind
\begin{eqnarray}\label{b3}
{\dot{\rho}_i} &=& F_i(\rho_1,\dots,\rho_n);
\nonumber \\
 F_i(\rho_1,\dots,\rho_n) &=& r_i^0 \rho_i \bigg \{ 1 -\sum_{j=1}^n
 (\alpha_{ij}^0 - r_j) \rho_j - \nonumber \\
 && \sum_{j=1}^n 
 \sum_{l=1}^n \alpha_{ij}^0 (\alpha_{ijl} + r_{il}) \rho_j \rho_l -
 \nonumber \\
 && \sum_{j=1}^n \sum_{k=1}^n \sum_{l=1}^n \alpha_{ij}^0 r_{ik} \alpha_{ijl}
 \rho_j \rho_k \rho_l \bigg \}.  \nonumber \\
\end{eqnarray}
We note that the system (\ref{b3}) consists of nonlinear ordinary differential 
equations with polynomial nonlinearities up to the order 4.
\section{Model equations when the birth rates are influenced by environmental
fluctuations}
Let us now suppose that the birth rates and interaction coefficients
depend on the density of the populations and in addition the birth rates
fluctuate. If the number of the populations in the studied system is $n$ then
in general the number of external influences that we have to account for will 
be $n$ too. The equations for the growth rates and interaction coefficients 
become 
\begin{eqnarray}\label{b4}
r_i &=& r_i^0 \left( 1 + \sum_{j=1}^n r_{ij} \rho_j\right) + \eta_i; 
\nonumber\\
\alpha_{ij} &=& \alpha_{ij}^0 \left(1+ \sum_{j=1}^n \alpha_{ijk} \rho_k \right).
\end{eqnarray} 
In Eq. (\ref{b4}) $r_{ij}$ and $\alpha_{ijk}$ are parameters and $\eta_i$
are  noises (Below we shall assume that $\eta_i$ are Gaussian white noises
but in general there is no restriction on the probability density function and
on the correlation properties of the noises).
\par
The substitution of Eq.(\ref{b4}) in Eq.(\ref{b1}) leads to a system
of model equations of the kind
\begin{eqnarray}\label{b5}
{\dot{\rho}_i} &=& F_i(\rho_1,\dots,\rho_n) +  \eta_i G_i(\rho_1,\dots,\rho_n);
\nonumber \\
 F_i(\rho_1,\dots,\rho_n) &=& r_i^0 \rho_i \bigg \{ 1 -\sum_{j=1}^n
 (\alpha_{ij}^0 - r_j) \rho_j - \nonumber \\
 && \sum_{j=1}^n 
 \sum_{l=1}^n \alpha_{ij}^0 (\alpha_{ijl} + r_{il}) \rho_j \rho_l -
 \nonumber \\
 && \sum_{j=1}^n \sum_{k=1}^n \sum_{l=1}^n \alpha_{ij}^0 r_{ik} \alpha_{ijl}
 \rho_j \rho_k \rho_l \bigg \}  \nonumber \\
 G_i(\rho_1,\dots,\rho_n) &=& \rho_i \bigg(1 - \sum_{j=1}^n \alpha_{ij}^0 \rho_j
 - \nonumber \\
 && \sum_{j=1}^n \sum_{k=1}^n \alpha_{ij}^0 \alpha_{ijk} \rho_j \rho_k \bigg).
 \nonumber \\ 
\end{eqnarray}
Thus the presence of noise in the growth rates leads to change of the kind
of the system of model equations. The system of nonlinear ordinary 
deterministic differential equations (\ref{b2}) is converted to a system of 
nonlieaar stochastic differential equations (\ref{b5}). In addition the 
stochastic terms $\eta_i G_i (\rho_1,\dots,\rho_n)$ in  the sytem (\ref{b5}) 
depend on the state of the system. The additive noise from Eqs.(\ref{b3})
leads to multiplicative noise in the system of equations (\ref{b5}). If all 
$\eta_i$ are Gaussian white noises then the system (\ref{b5}) can be converted 
to a Fokker-Planck equation for the probability density function of the 
densities of the populations.
\section{Model equations when the interaction coefficients are influenced by
environmental fluctuations}
This case is more complicated as the number of interagtion coefficients in
general is $n^2$ where $n$ is the number of interacting populations.
The additive noises $\sigma_{ij}$ are included in the equations for $\alpha_{ij}$
\begin{eqnarray}\label{b6}
r_i &=& r_i^0 \left( 1 + \sum_{j=1}^n r_{ij} \rho_j\right) ; 
\nonumber\\
\alpha_{ij} &=& \alpha_{ij}^0 \left(1+ \sum_{j=1}^n \alpha_{ijk} \rho_k \right)
+ \sigma_{ij}.
\end{eqnarray} 
In Eq. (\ref{b6}) $r_{ij}$ and $\alpha_{ijk}$ are parameters and $\eta_i$
are Gaussian white noises.
\par
The substitution of Eq.(\ref{b6}) in Eq.(\ref{b1}) leads to a system
of model equations of the kind
\begin{eqnarray}\label{b7}
{\dot{\rho}_i} &=& F_i(\rho_1,\dots,\rho_n) - \nonumber \\
&& \sum_{j=1}^n \sigma_{ij} G_{ij}(\rho_1,\dots,\rho_n);
\nonumber \\
 F_i(\rho_1,\dots,\rho_n) &=& r_i^0 \rho_i \bigg \{ 1 -\sum_{j=1}^n
 (\alpha_{ij}^0 - r_j) \rho_j - \nonumber \\
 && \sum_{j=1}^n 
 \sum_{l=1}^n \alpha_{ij}^0 (\alpha_{ijl} + r_{il}) \rho_j \rho_l -
 \nonumber \\
 && \sum_{j=1}^n \sum_{k=1}^n \sum_{l=1}^n \alpha_{ij}^0 r_{ik} \alpha_{ijl}
 \rho_j \rho_k \rho_l \bigg \}  \nonumber \\
 G_{ij}(\rho_1,\dots,\rho_n) &=& \rho_i \rho_j r_i^0  \bigg(1 + \sum_{k=1}^n
 r_{ik}^0 \rho_k \bigg).
 \nonumber \\
\end{eqnarray}
In the general case the system (\ref{b7}) can be solved only numerically.
But in the particular cases (where each equations contains only a single 
multiplicative noise and this multiplicative noise is Gaussian white noise)
the analytical treatment is possible on the basis of the theory
of Markov processes and forward Kolmogorov (Fokker-Planck) equation. 
\section{Model equations for the general case when all parameters are influenced
by environmental fluctuations}
In the general case the environment fluctuations can influence both the
growth rates and the interaction coefficients. 
In this case the additive noises $\sigma_{ij}$ are present in the equations for 
$\alpha_{ij}$ and additive noises $\eta_i$ are present in the equation for $r_i$.
Thus the equations for the growth rates and for the competition coefficients
become
\begin{eqnarray}\label{b8}
r_i &=& r_i^0 \left( 1 + \sum_{j=1}^n r_{ij} \rho_j\right) + \eta_i ; 
\nonumber\\
\alpha_{ij} &=& \alpha_{ij}^0 \left(1+ \sum_{j=1}^n \alpha_{ijk} \rho_k \right)
+ \sigma_{ij}
\end{eqnarray} 
The substitution of Eqs.(\ref{b8}) in Eq.(\ref{b1}) leads to a system
of model equations of the kind
\begin{eqnarray}\label{b9}
{\dot{\rho}_i} &=& F_i(\rho_1,\dots,\rho_n) - \nonumber \\
&& \sum_{j=1}^n \sigma_{ij} 
G^{(1)}_{ij}(\rho_1,\dots,\rho_n) \nonumber \\
&&  + \eta_i G^{(2)}_i - G^{(3)}_{i};
\nonumber \\
 F_i(\rho_1,\dots,\rho_n) &=& r_i^0 \rho_i \bigg \{ 1 -\sum_{j=1}^n
 (\alpha_{ij}^0 - r_j) \rho_j - \nonumber \\
 && \sum_{j=1}^n 
 \sum_{l=1}^n \alpha_{ij}^0 (\alpha_{ijl} + r_{il}) \rho_j \rho_l -
 \nonumber \\
 && \sum_{j=1}^n \sum_{k=1}^n \sum_{l=1}^n \alpha_{ij}^0 r_{ik} \alpha_{ijl}
 \rho_j \rho_k \rho_l \bigg \}  \nonumber \\
 G^{(1)}_{ij}(\rho_1,\dots,\rho_n) &=& \rho_i \rho_j r_i^0  \bigg(1 + \sum_{k=1}^n
 r_{ik}^0 \rho_k \bigg)
 \nonumber \\
 G^{(2)}_i &=& \rho_i \bigg(1 - \sum_{j=1}^n \alpha_{ij}^0 \rho_j
 - \nonumber \\
 && \sum_{j=1}^n \sum_{k=1}^n \alpha_{ij}^0 \alpha_{ijk} \rho_j \rho_k \bigg)
 \nonumber \\ 
 G^{(3)}_i &=& \eta_i \rho_i \sum_{j=1}^n \sigma_{ij} \rho_j.
\end{eqnarray}
We observe three kinds of noise terms in the system of equations
(\ref{b9}). $G^{(1)}_{ij}$ is a result of the action of the environment on the
coefficients of interaction between the populations. $G^({2})_i$ is a result
of the action of the environment on the growth rates. And because of the
nonlinearity of the model equations there exist third kind of terms $G^{(3)}_i$
that is a result of the joint action of the two influences. If one kind of 
influence is not present $G^{(3)}$ is $0$. In general the system (\ref{b9})
can be studied only numerically. Analytical treatment is possible only when one
of the two kinds of influences is missing and the noises that account for the
environment influences are Gaussian white noises. 
\section{General case for presence of multiplicative white noise in the
coefficients}
Even more general case of influence by the environment is when this
influence depends on the state of the system. In this case instead of additive
noises we have to add multiplicative noises at the equations for the
growth rates and interaction coefficients. The equations become
\begin{eqnarray}\label{b10}
r_i &=& r_i^0 \left( 1 + \sum_{j=1}^n r_{ij} \rho_j\right) + \eta_i H_i
(\rho_1,\dots, \rho_n) ; 
\nonumber\\
\alpha_{ij} &=& \alpha_{ij}^0 \left(1+ \sum_{j=1}^n \alpha_{ijk} \rho_k \right)
+ \sigma_{ij} I_{ij}(\rho_1, \dots,\rho_n). \nonumber \\
\end{eqnarray} 
We remember that
in Eq. (\ref{b10}) $r_{ij}$ and $\alpha_{ijk}$ are parameters; $\eta_i$
and $\sigma_{ij}$ are Gaussian white noises; and $H_i$ and $I_{ij}$ are 
functions depending on the densities of the populations.
The substitution of Eq.(\ref{b10}) in Eq.(\ref{b1}) leads to a system
of model equations of the kind
\begin{eqnarray}\label{b11}
{\dot{\rho}_i} &=& F_i(\rho_1,\dots,\rho_n) - \nonumber \\
&& \sum_{j=1}^n \sigma_{ij}I_{ij}(\rho_1,\dots,\rho_n) \times \nonumber \\
&& G^{(1)}_{ij}(\rho_1,\dots,\rho_n) \nonumber \\
&&  + \eta_i H_i(\rho_1,\dots,\rho_n) G^{(2)}_i - G^{(3)}_{i};
\nonumber \\
 F_i(\rho_1,\dots,\rho_n) &=& r_i^0 \rho_i \bigg \{ 1 -\sum_{j=1}^n
 (\alpha_{ij}^0 - r_j) \rho_j - \nonumber \\
 && \sum_{j=1}^n 
 \sum_{l=1}^n \alpha_{ij}^0 (\alpha_{ijl} + r_{il}) \rho_j \rho_l -
 \nonumber \\
 && \sum_{j=1}^n \sum_{k=1}^n \sum_{l=1}^n \alpha_{ij}^0 r_{ik} \alpha_{ijl}
 \rho_j \rho_k \rho_l \bigg \}  \nonumber \\
 G^{(1)}_{ij}(\rho_1,\dots,\rho_n) &=& \rho_i \rho_j r_i^0  \bigg(1 + \sum_{k=1}^n
 r_{ik}^0 \rho_k \bigg)
 \nonumber \\
 G^{(2)}_i &=& \rho_i \bigg(1 - \sum_{j=1}^n \alpha_{ij}^0 \rho_j
 - \nonumber \\
 && \sum_{j=1}^n \sum_{k=1}^n \alpha_{ij}^0 \alpha_{ijk} \rho_j \rho_k \bigg)
 \nonumber \\ 
 G^{(3)}_i &=& \eta_i H_i(\rho_1,\dots,\rho_n) \rho_i \sum_{j=1}^n \sigma_{ij}
 \times \nonumber \\
&& I_{ij}(\rho_1,\dots,\rho_n) \rho_j.  \nonumber \\
\end{eqnarray}
\section{An example for reduction of  model stochastic differential
equations to a Fokker-Planck equation}
Let us discuss the model system (\ref{b5})
for the case of one population (we set $r^0=r$; $r_{11}=0$; $\alpha_{11}^0
=\alpha$; $\alpha_{111}=0$). The model equation is
\begin{eqnarray}\label{b12}
\dot{\rho} &=& F(\rho) + \eta G(\rho) \nonumber \\
F(\rho)&=& r \rho - \alpha r \rho^2; \ \ G(\rho) = \rho - \alpha \rho^2 .
\end{eqnarray}
\par
Eq.(\ref{b12}) is a particular case of a more general equation. 
We shall discuss the case where $F(\rho)$ and $G(\rho)$
are polynomials of arbitrary orders $p_1$ and $p_2$, i.e.,
\begin{equation}\label{b13}
F(\rho)= \sum_{i=1}^{p_1} \mu_i \rho^i; \ \ G(\rho) = \sum_{i=1}^{p_2} 
\theta_i \rho^i.
\end{equation}
where $\mu_i$ and $\theta_i$ are parameters.
In this case Eq.(\ref{b12}) becomes
\begin{eqnarray}\label{b14}
\dot{\rho} &=& \sum_{i=1}^{p_1} \mu_i \rho^i + 
\eta \sum_{i=1}^{p_2} 
\theta_i \rho^i . \nonumber \\
\end{eqnarray}
The formal integration of Eq.(\ref{b14})  leads to the equation
\begin{equation}\label{b15}
\rho (t) = \rho(t=0) + \int_0^t d \tau \ F[\rho(\tau)] + 
\int_0^t dW_\tau \ \ G[\rho(\tau)],
\end{equation}
where $W_\tau$ is a Wiener process.
The integral $\int\limits_0^t dW_\tau \ \ G(\rho(\tau))$ can be integral of Ito 
kind or integral of Stratonovich kind. Let us assume that the integral is an 
integral of Ito kind.  For this case Eq.(\ref{b15}) can be written as
\begin{equation}\label{b16}
d\rho_t =  F(\rho_t)dt +  G(\rho_t) dW_t,
\end{equation}
where we wrote the time dependence as subscript and in general $F$ and $G$
are given by Eqs.(\ref{b13}). The Fokker-Planck equation
that corresponds to Eq.(\ref{b16}) is
\begin{eqnarray}\label{b17}
\frac{\partial }{\partial t}p(x,t) = - \frac{\partial}{\partial x} \Bigg \{
p(x,t) \bigg[ \sum_{i=1}^{p_1} \mu_i x^i \bigg] \Bigg \} + \nonumber \\
\frac{1}{2} \frac{\partial^2}{\partial x^2} \Bigg \{p(x,t) \bigg[ 
\sum_{i=1}^{p_2} \sum_{j=1}^{p_2} \theta_i \theta_j x^{i+j} \bigg] \Bigg \}.
\nonumber \\
\end{eqnarray}
We can formulate the following statement (the proof will be presented 
elsewhere): \emph{
Let $b_1$ and $b_2$ be natural boundary points ($-\infty \le b_1 < b_2 \le 
\infty $). Let in addition $\sigma(x) = \sum\limits_{i=1}^{p_2} \theta_i x^i >0$ in $(b_1,b_2)$. Then
the diffusion process $X_t$ that is solution of the stochastic differential
equation Eq.(\ref{b16}) has unique invariant distribution with
p.d.f.
\begin{eqnarray}\label{b18}
p^0(x) = \nonumber \\
\frac{\cal{N}}{\sum\limits_{i=1}^{p_2} \sum\limits_{j=1}^{p_2} 
\theta_i \theta_j x^{i+j}} \exp \left(  \int_c^x dy 
\frac{2 \sum\limits_{i=1}^{p_1} \mu_i y^i }{\sum\limits_{i=1}^{p_2}
\sum\limits_{j=1}^{p_2} \theta_i \theta_j y^{i+j}} \right), \nonumber \\ 
\vee x \in (b_1,b_2)
\end{eqnarray}
if the quantity
\begin{eqnarray}\label{b19}
{\cal{N}}^{-1}= \nonumber \\
\int_{b_1}^{b_2} dx \frac{1}{\sum\limits_{i=1}^{p_2} 
\sum\limits_{j=1}^{p_2} \theta_i \theta_j x^{i+j}} \exp \left( \int_c^x 
dy \frac{2 \sum\limits_{i=1}^{p_1} \mu_i y^i}{\sum\limits_{i=1}^{p_2}
\sum\limits_{j=1}^{p_2} \theta_i \theta_j y^{i+j}} \right), \nonumber \\
 b_1 < c < b_2
\end{eqnarray}
has finite value. In addition each time-dependent solution $p(x,t)$ of the
Fokker-Planck equation (\ref{b17}) in $(b_1,b_2)$ satisfies
\begin{equation}\label{b20}
\lim_{t \to \infty} p(x,t) = p^0(x)
\end{equation}
}
\par
For the case of more than one population
we have to solve the system of stochastic differential equations
\begin{eqnarray}\label{c8}
dX_i(t) = F_i[X_1(t),\dots,X_n(t)] + \nonumber\\
 G_{i}[X_1(t),
\dots,X_n(t)] dW_i(t), \ i=1,\dots,n,
\end{eqnarray}
where $W_j(t)$ are independent Wiener processes and 
\begin{eqnarray}\label{c9}
 F_i(\rho_1,\dots,\rho_n) &=& r_i^0 \rho_i \bigg \{ 1 -\sum_{j=1}^n
 (\alpha_{ij}^0 - r_j) \rho_j - \nonumber \\
&& \sum_{j=1}^n 
 \sum_{l=1}^n \alpha_{ij}^0 (\alpha_{ijl} + r_{il}) \rho_j \rho_l -
 \nonumber \\
 && \sum_{j=1}^n \sum_{k=1}^n \sum_{l=1}^n \alpha_{ij}^0 r_{ik} \alpha_{ijl}
 \rho_j \rho_k \rho_l \bigg \}  \nonumber \\
 G_i(\rho_1,\dots,\rho_n) &=& \rho_i \bigg(1 - \sum_{j=1}^n \alpha_{ij}^0 \rho_j
 - \nonumber \\
&& \sum_{j=1}^n \sum_{k=1}^n \alpha_{ij}^0 \alpha_{ijk} \rho_j \rho_k \bigg). 
\end{eqnarray}
The corresponding Fokker-Planck equation is: ($G_{ij}=G_i \delta_{ij}$ where
$\delta_{ij}$ is the Kronecker delta-symbol)
\begin{eqnarray}\label{c10}
\frac{\partial}{\partial t} p = - \sum_{i=1} \frac{\partial}{\partial x_i}
[p F_i (x_1,\dots,x_n,t)] + \nonumber \\ \frac{1}{2} \sum_{i=1}^n \sum_{j=1}^m
\frac{\partial}{\partial x_i} \frac{\partial}{\partial x_j}[p
\times \nonumber \\
G_{ij}(x_1,\dots,x_n,t)G_{ji}(x_1,\dots,x_n,t)].
\end{eqnarray}
\section{Concluding remarks}
In this paper we discuss the influence of environment fluctuations on the
dynamics of interacting populations modelled by system of nonlinear differential
equations. The problem is intereting as the fluctuations  are often present 
in the complex systems \cite{f1} -\cite{kv} and in particular in the systems 
of populations \cite{t1}-\cite{t3}.
\par
In the discussed above models the growth rates and the interaction 
coefficents depend on the density of the populations. The influence of environment leaded to terms
containing multiplicative noise or more complicated kind of noise. This research
is continuation of our research on presence of additive noise in the model equations
of the population dynamics \cite{biom1},\cite{k1} 
There are two main approaches to deal with fluctuations. The first approach
is based on appropriate averaging and in this way one investigates mean
quantities connected to the problem. In addition this approach can 
lead to reduction of the spatial dimensions of the problem if such
dimensions are present as well as it can lead to relatively simple mathematical
description of somlex media such as porous media (for examples see
\cite{ott1,ott2}).
\par
Finally let us note that one possible  extension of the above research 
is to include  spatial dimensions in the model equations \cite{vit09}-\cite{vit10a}.
For the case without environment influence we can obtain even analytical solutions of the
model nonlinear PDEs \cite{vdk10}-\cite{vit09x}. In general the case when
environment influence is present can be treated only numerically. We
shall discuss these problems in more detail elsewhere.

\end{document}